\documentclass[graybox]{svmult}

\usepackage{type1cm}        
\usepackage{makeidx}         
\usepackage{graphicx}        
\usepackage{sidecap}          
           
\usepackage{multicol}        
\usepackage[bottom]{footmisc}

\usepackage{url}

\usepackage{newtxtext}       %
\usepackage{newtxmath}       

\usepackage{tikz}
\usepackage{mathtools}

\makeindex             
                       
\newcommand{\be}{\nopagebreak[3]\begin{equation}}
\newcommand{\ee}{\end{equation}}
\newcommand{\ba}{\nopagebreak[3]\begin{eqnarray}}
\newcommand{\ea}{\end{eqnarray}}

\newcommand{\bc}{}

\begin{document}

\title*{Time, space and matter \\ in the primordial universe}
  \titlerunning{Time, space and matter in the primordial universe} 
\author{Francesca Vidotto}
\institute{Francesca Vidotto \at The University of Western Ontario\\ \emph{Department\,of\,Physics\,and\,Astronomy, Department\,of\,Philosophy, Rotman\,Institute\,of\,Philosophy}\\ 
Western University is located in the traditional lands of Anishinaabek, Haudenosaunee, L\=unaap\'eewak and Attawandaron peoples.\\
\email{fvidotto@uwo.ca}\\[1em]
{\it This is a preprint of the following chapter: Francesca Vidotto, Time, space and matter in the primordial universe, to be published in ``Advances in Cosmology'', edited by Streit-Bianchi Marilena, Catapano Paola, Galbiati Cristiano, Magnani Enrico, 2022, Springer, reproduced with permission of Springer. 
}
}
%
%
\maketitle

\abstract{Time, space, and matter are categories of our reasoning, whose properties appear to be fundamental. However, these require a scrutiny as in the extreme regime of the primordial universe these present quantum properties. What does it mean for time to be quantum? What does it mean for space? Are space and time disappearing, or what is disappearing are simply the categories we have been using to understand them? Concepts such as the superposition of causal structures or the quantum granularity of space require our attention and should be clarified to understand the physics of the primordial universe. The novelty that this brings requires us to reflect on matter as well: How can matter be defined on a granular space? Is quantum gravity hinting us toward considering new types of matter? The answers to these questions, that touch the foundations of physics and the very concepts with which we organize our understanding of reality, require in the end of the journey to confront ourselves with empirical data. And for that, the universe itself provides us with the best of possible laboratories.}


\section{Introduction}

General Relativity and Quantum Mechanics are the two great revolutions of XX century physics. Both have changed our perspective on the universe. Both have expanded the boundaries of the universe, one towards large scales, the other towards small scales. They are  conceptual revolutions in the way we understand space and time, and matter, respectively.  Shaking the very foundations of physics,  these two revolutions have opened the possibility of promoting our knowledge of the universe as a whole to an exact science. The birth of modern cosmology is inextricably linked to what we have understood in relativity and quantum theory.

A remarkable aspect of modern cosmology is precisely that it needs to combine our knowledge about small scales and our knowledge about large scales. Only this convergence can give us a coherent vision of the universe. Convergence and coherence reinforce our confidence in the physical theories we use. In this sense, cosmology is a science that brings everything together: all the physics we know. 

This does not mean that cosmology is the science that describes everything, namely everything the universe contains \cite{Vidotto:2015bza}. On the contrary, cosmology was born understanding that it is possible to describe the universe at its largest accessible scale with simple laws, by treating it as a whole and leaving out the details. The  embodiment of this idea is Einstein's ``cosmological principle'' \cite{Einstein:1917}. This principle allows to apply Einstein's theory of gravitation to the whole universe. This is an application of General Relativity where we can use exact solutions.  The idea is simple and ingenious: it is possible to capture the dynamics of the universe using a single quantity, for example the radius or the volume. More precisely, it is possible to reduce the infinite number of degrees of freedom of the gravitational field down to a model with a single degree of freedom. This is enough to capture the large scale dynamics of the universe, that is, how the universe evolves as a whole.

This is what Einstein discovered in 1917: it is possible to write a single equation to describe the large scale dynamics of the universe.  This equation shows that the universe must evolve, but Einstein resisted what his own equations were indicating, and it was only later,  thanks to Georges Lema\^itre \cite{Lemaitre:1927}, that it was realized that indeed the universe could not be static but had to expand. On the other hand this corresponds to the heart of Einstein's General Relativity: space-time is identified with the gravitational field \cite{Einstein:1952}, and this, like all fields, has a dynamic nature.

The gravitational field depends on the presence of matter. The physical approximation that Einstein takes, with the cosmological principle, consists in assuming that, on large scale, matter can be considered as uniformly distributed. Einstein imagines that the matter uniformly distributed are stars that fill the whole universe. Today we know that Einstein's assumption is valid, but for the very large scale distribution of galaxies, not stars. In fact, it is only almost 10 years after Einstein writes his equations that it is understood that nebulae are other galaxies \cite{Hubble:1925}: the universe is far bigger than we thought. The step taken by Einstein  introducing the cosmological principle can be interpreted in different ways. It is an approximation, which is not valid  within our galaxy or  with respect to the local group of galaxies of which the Milky Way is a part. But it is an approximation that works  well on the largest scale, where the distribution of matter we consider is given by the average distribution of galaxy clusters.

It is largely a philosophical motivation that drives Einstein to take the cosmological principle as the basis of his cosmology \cite{Smeenk:2012}. The cosmological principle tells us that there is no privileged observer in the universe, hence all observers see the same universe. This  leads to a homogeneous and isotropic universe.  This is a reinforced version of the Copernican principle, according to which our position on Earth is not a privileged one. Einstein is guided by the philosophical ideas of Ernst Mach \cite{Holton:1968}. According to Mach, the inertia of bodies should be accounted by the distribution of matter in the universe.  Einstein had arrived at the  the equations of General Relativity and the idea that physics should not depend on the choice of coordinates, under the influence of the ideas of Mach.  He generalizes the idea that there are no privileged points in the universe, to the idea that there are no privileged reference systems of the universe. This way of looking at things proves extremely fruitful. It leads to a simplification of our vision of the world in which there are only general covariant substances: such must be the nature at the fundamental level of quantum fields as well as of the universe itself.

In the rest of the article, I discuss this understanding about time, space, and matter, based on what we know from General Relativity and quantum theory, particularly how this affects our understanding of the universe.

\section{Time}
The central idea of General Relativity is the identification of space-time with the gravitational field.  Distances (that in non relativistic physics are given by the cartesian coordinates {\it x, y, z}\,) and durations (that in non relativistic physics are given by the cartesian coordinate {\it t}\,) depend on the gravitational field.  Hence measurable physical time is a dynamical variable, on the same level as the other physical variables. This fact is one of the most important and disconcerting aspects of General Relativity. Special Relativity already relates space and time, conceptually joining them into a single entity, so that it is no longer possible to think of time in absolute terms as in Newtonian mechanics.  But in General Relativity there is a far more radical step: frames of reference are local, they cannot be extended to the whole universe, and can only be valid in a limited region of space time. In particular, we have to associate a different time with every local frame of reference: time becomes multi-fingered. Instead of viewing it as a single abstract quantity, we have to consider it as the local reading of physical quantities that we take as {\em clocks}.

From a formal point of view, what happens in General Relativity, is that in the Hamiltonian formalism the evolution is coded in a set of constraints \cite{Rovelli:2021}.  The coordinate time (as opposed to the physical time measured by clocks) is a dispensable gauge dependent quantity.  It is often said in General Relativity or in quantum gravity that time disappears from the equations.  This does not mean that these theories do not describe evolution; it  means that the theory describes the evolution with respect to a dynamic variable that is identified as the clock.  There is nothing mysterious about this, yet the disappearance of time has confused many scientists, also in the development of cosmological models. In cosmology, it is possible to choose a privileged global clock by exploiting the homogeneity and isotropy of the universe on a large scale. This is  a consequence of the specific approximations made in classical cosmology. In this way in cosmology it is possible to choose a preferred time with respect to which to write the evolution equations.  But this ceases to be true when we promote these same equations to quantum equations: it is no longer possible to choose the privileged time of cosmology and we must return to a writing of the equations in terms of constraints \cite{Rovelli:2015gwa}.

In the early universe, quantum effects on the dynamics of the gravitational field cannot be disregarded. To describe them we need the quantum version of the equations of the gravitational field. We know that all physical fields are quantum fields, but while we have equations that describe the quantum behavior of the fields of matter well, the quantum equations of the gravitational field have not yet found a definitive form. The most famous equation of quantum gravity is the Wheeler-deWitt equation \cite{WdW}. When it was written, this equation puzzled physicists because the evolution it describes does not select any privileged time, so that this equation has become the prototype of a timeless equation. Half a century later, we now understand that it describes evolution is with respect to any dynamic variable, a clock, which is itself quantum.

The Wheeler-deWitt equation was not well defined and it is not easy to handle. It was initially studied in the simplified context of cosmology, where there is only one degree of freedom to evolve. The object that a physicist faces at this point is breathtaking: the equation describes a quantum universe evolving in quantum time. What may this mean?

A natural objection to a quantum universe is the idea that Quantum Mechanics applies only to small scales, and not to huge systems such as the universe. But this is misleading for two reasons. First, Quantum Mechanics does not apply only to small objects, quantum effects depend on size, energy and other variables. Second, the universe is expanding and it was very small in the early stages of this expansion.  In fact, if we extrapolate backwards the history of the universe, we need to imagine that all the matter in the universe was contained in the size of a proton. This implies an energy density for which there is no doubt that the universe behaves quantumly.  This conclusion forces us to seriously consider the realm of quantum cosmology.

A characteristic phenomenon of quantum physics is superposition. When this is applied to macroscopic objects, quantum superposition generates the conceptual confusions illustrated by Schr\"odinger's famous cat which is in a superposition of two macroscopically different states. That is, its classical state is not defined.  The way to clarify this confusion is still debated and affects our way of thinking of the early universe.  In the case of a superimposed universe, we must think that the early universe was such that its geometry and its physical time were not defined. The fact that the universe is expanding today is encoded in its geometry; the time reversed geometry would be a universe that is contracting instead. In the early universe these two geometries may have been in a superposition. In the same way, it is possible for temporal directions to be in superposition. This makes it possible to switch from one geometry to another, from one time to another. This is a typical effect of Quantum Mechanics: tunnelling. From the point of view of the evolution of the universe, this allows for a universe that first contracts, then bounces and begins to expand \cite{Ashtekar:2007tv}.

The effects of quantum superposition on the structure of time are not confined to cosmology.   Even in elementary non relativistic Quantum Mechanics it is possible to have the quantum superposition of two histories in which the temporal order in which to events happen is different.  These are called indefinite causal structures \cite{Castro-Ruiz:2017gzs}. In this sense "quantum time" is much closer to us than it might seem if confined to the early universe.

In cosmology, this superposition of causal structures generates another sense for which time disappears in quantum cosmology, distinct from the general relativistic multiplicity of clock times.   If a clock time does not determine a unique causal structure, it looses its specificity as a temporal variable. Temporal and spacial directions get mixed up.   In a universe with four dimensions all four of these dimensions are then on the same footing and the resulting geometry is no longer that of the space-time considered by Einstein (the space of Minkowski and Lorentz) but rather a space closer to Euclidean geometry in four dimensions.  In a celebrated work \cite{Hartle:1983ai},  Stephen Hawking and James Hartle suggested that the early universe could have emerged form such an "Euclidean" phase, where the spacial and temporal directions of spacetime are no longer distinguished.  Their suggestion for the early quantum universe is known as ``no-boundary'' proposal.  In this model, if we go back to the early universe, we reach a point where it is no longer possible to go further back because time disappears. This is the picture that Hawking illustrated in his famous book ``A Brief History of Time.'' 

The issue, however, is open, and much debated. It is possible that the ``no-boundary'' proposal captures an aspect of time in the early universe: the lack of a specific time structure. But it might also be that what this proposal captures only a piece of a story in which the universe transitions from a contracting phase to an expanding one.  Also in this case there is ``no-boundary", that is there is no beginning of time with which one collides by going further and further back in time. But in this case the universe may have had its own evolution before the Euclidean phase considered by Hartle and Hawking.

A further clarification is important. In the discussion so far, we have discussed causal structure. This is encoded in the metric, the mathematical object that contains the information on the gravitational field and on the geometry. When in the early universe geometry is transformed into Euclidean geometry, this is equivalent to a change in the signature of the metric. But caution: this causal structure captures the distinction between space and time direction in spacetime, it does not capture the {\em direction} of time. The direction  of time is a phenomenon that is not described by General Relativity. Fields, including the gravitational one, are described by mechanical laws, be they classical or quantum.  Mechanics, including Quantum Mechanics, is invariant under CPT, namely inversion of charge, parity and time inversions. Hence the dynamics does not have a privileged direction of time.  Instead, the directionality of time is a characteristic of thermodynamic phenomenon. In the absence of heat sources, a hot body becomes cold; a very orderly system becomes disordered over time; a system that can be described with one or a few degrees of freedom will evolve into a system that requires many degrees of freedom to be described: in a single concept, observed entropy is always lower in the past and higher in the future. It is this thermodynamic concept of entropy that contains information on the direction of time.

In our universe we observe entropy to uniformly increase. Whatever the entropy of the universe today, this entropy must have been much lower in the past. This fact is known and discussed under the name of  ``Past Hypothesis'' \cite{Albert:2000}.  Since low entropy can be thought as a "special" condition, this fact raises the question why was the universe in special initial conditions in the past? The discussion is open and involves physicists and philosophers in equal measure. We do not have an answer yet, but my feeling is that we must continue on the path that General Relativity and Quantum Mechanics have indicated to us up to now: we must not think in absolute terms, but in relational terms. Perhaps this is teaching us, once again that we should not think of the conditions of the universe as special at first, but that this is an effect of our point of view and our particular way of interacting with matter and space-time.

\section{Space}

Einstein's relativity links space and time together, and space-time to matter. This link plays a fundamental role in the early universe for the formation of structures. The universe is homogeneous on a very large scale but it is not at other scales: gravity generates an instability in the homogeneity: any excess of matter is a source of gravitational attraction and  becomes a seed for the formation of structures such as stars, black holes, galaxies, galaxy clusters, and so on. In the early universe there must therefore have been small deviations from global homogeneity. These inhomogeneities, which we can think of as deviations from a homogeneous distribution of matter, are now understood to be of quantum origin. When the universe comes out of its primordial quantum phase and becomes classical, the quantum fluctuations of matter turn into statistical fluctuations in the distribution of matter. The fact that matter has a quantum nature appears therefore to be an essential element in the explanation of the structures that populate the universe.

In Einstein's equations, space-time and matter are interconnected. The dynamics of space-time is determined by the presence of matter, and vice versa, matter moves according to the shape of space-time. Consequently, if matter has quantum properties, this link implies that geometry must have them too. What does it mean for spacetime to be quantized? Although to fully answer this question it is necessary to understand the quantization of whole space-time, it is easier to understand and explain it by focusing on space: let's fix an instant in time, and consider  quantized space.

A characteristic phenomenon of quantum physics is the possibility for a system to be in a superposition of different states. This introduces a form of fuzziness or indeterminacy, into all quantum systems: physical variable can fail to have a determined value. We have seen how this can concretely manifest in the case of the geometry of space-time: in a great rebound, the universe can be in a state where it is simultaneously contracting and expanding. The superimposition of geometries, which plays a role in the description of the universe at very large scale, as well as on the scale of possible laboratory experiments too, becomes more and more important as we approach shorter scales, until it touches the smallest scale of all, that of Planck \cite{Rovelli:2014ssa}.

This leads us to another aspect of quantization, the most fundamental, together with superposition, which gives the theory its name: everything manifests itself in the form of discrete quantities, or "quanta". Let us remember the case of the electromagnetic field: the continuous field can be thought of as emerging from a set of photons, the quanta of the electromagnetic field. Similarly, in gravity, space-time can be thought of as emerging from a set of quanta of the gravitational field. It is necessary to be cautious in the way in which we speak of emergency in this context: both in the electromagnetic case and in the gravitational case, the emergence has the clear meaning of a classical limit. In other words, this is the familiar process by which we loose properties characteristic of the quantum regime, which are replaced by other characteristics of the classical regime. In particular, in the case of fields, we pass from a discrete description to a continuous one.

Can we say that space disappears in the early universe when the gravitational field enters a quantum regime? If we identify space with its property of continuity, then yes, it disappears. But the continuity property is not an essential property for the gravitational field. The continuity property is useful for describing contiguity and locality of processes at scales much larger than the discreteness scale. What remains of continuity and locality, in a quantized space? The information at the basis of continuity and locality must be rooted in some properties of the quantum states of space. In a sense, this is what guides us in their definition. The quantum states of the gravitational field, at a given time, must encode the information on which quantum of space is adjacent to which other, and which quantum is interacting with each other. In the case of Loop Quantum Gravity \cite{Gambini:2020ypi,Gambini:2011zz} this idea is embodied in quantum networks (spinnetworks) which encode the contiguity relationships between the different quanta of space. The interactions between the quanta of space constitute a network which, in the classical limit, results in the familiar space described by General Relativity.

To what extent does each quantum space contain properties that correspond to those we attribute to classical space? Single quanta of space embody a notion of geometry (based on Roger Penrose's spin-geometry theorem \cite{Penrose:1966}) and therefore a notion of distance, the key element in building the theory of General Relativity. Spin networks can thus be associated  to a metric, but it will be a fuzzy metric. A quantum of space can also have information on the number of dimensions and on the type of geometry (Lorentzian or Euclidean). Just as a photon is already a manifestation of the electromagnetic field even when the classical limit of the theory is not considered, in the same way a quantum of space is in many respects a manifestation of the gravitational field, even when its classical regime is not considered.

An intuitive way of thinking about the quanta of the gravitational field is that each corresponds to a small unit of space: let's take a certain region of space and divide it into many tiny pieces, but not smaller than the Planck scale. The Planck area and the Planck volume, represent in fact a minimal size for space. This is analogous to the fact that the energy of photons of a given frequency is proportional to the Planck constant. But, just as for photons we cannot simply think of the electromagnetic field just as a classical sum of small Planck energy packets, in the same way we cannot think that space is simply constituted by adding up many Planck volumes.

The quantization that manifests itself for the quanta of space is genuine, it is not an artifice of having taken a discretization of space. This means for example that for each individual quantum of space, if we measure its properties such as its area or its volume, we have the characteristic quantum leaps between one value and another: For example, between the zero value of the volume and the first smaller one measured value, i.e. a jump, and between the first and the second there is another jump.

Let me illustrate this fact with a specific example. In Quantum Mechanics the angular momentum is quantized: it can only take certain values, in jumps. Notice that the fact that the components of the angular momentum are quantized does not break the symmetry of the theory under rotations. In a rotated frame, the eigenstates are rotated, but the eigenvalues are the same.  Similarly, the fact that space is quantized does not break the characteristic symmetries of space and space-time - and in particular the Lorenz symmetry is preserved \cite{Rovelli:2010ed}: when I observe the quanta of space in a moving frame of reference, I still find the same jumps quantum in the area and in the volume, but with different probabilities.

In Quantum Mechanics, as energy rises, the intervals between high energy states, namely the quantum leaps at higher energy, become smaller and less visible. Similarly in gravity, as a larger space is considered, the more quantum leaps in area and volume become smaller and less visible. When we consider macroscopic regions of space, and even more so considering cosmological scales, the granularity of space is negligible. The quanta of space have a characteristic scale which is the Planck scale, a very minute scale: there is the same distance in terms of orders of magnitude between us and the scale of a proton, as between the scale of the proton and that of a quantum of space.

And yet, the granularity of space plays an important role in the early universe. When we study the quantum fluctuations from which every structure derives, these fluctuations have to do not only with matter but through Einstein's equation also with geometry. To understand the quantum fluctuations of geometry and to study how they evolve it is necessary to give a description that takes into account the granularity of space \cite{Ashtekar:2015dja,Gozzini:2019nbo}.

\section{Matter}

In the early universe, the essential feature of matter fields, namely the fact that they are quantized, needs to be extended to the gravitational field, which is also quantized. Conversely, from General Relativity, we have learnt another important property that a field must possess: it must not depend on the coordinates with which it is described. One way to express this is the fact that the theory is independent of background. The gravitational field provides a background to the fields of matter by choosing a fixed configuration, that is, by neglecting their dynamic nature. The presence of a background and the fact that this is continuous form the basis of the whole construction of the theory of fields, both classical and quantum. Describing matter at a fundamental level therefore means doing it in a way that respects general covariance and is compatible with the quantum discretion of space-time.

One way to achieve this is to think of spinnetwork states as lattices, albeit special lattices as they are invariant under diffeomorphisms, i.e. the coordinate transformations that characterize General Relativity. Precisely from the fundamental physics developed using lattices come some techniques at the basis of the quantization of gravity. One such method are Wilson loops, which are at the basis of Loop Quantum Gravity \cite{Rovelli:2004tv}. Wilson lines and loops are known objects in fundamental physics, especially for the study of strong interactions.

An interesting observation is that the loop quantization can be applied not only in the path-integral formalism but also in the standard Hamiltonian formulation of Quantum Mechanics, \'a la Dirac. In addition to the case of space quanta, we can use this same quantization technique for ordinary particles as well. It can be shown that the usual quantization is found within the loop quantization when the loops become very short, that is, in a regime where their length can be neglected. The usual quantization simply becomes then a special case of the general case of loop quantization.

The quantization of space-time in Loop Quantum Gravity is compatible with the Standard Model of particles and with its possible extensions. In this sense, quantum gravity presents no problems and provides a coherent picture in which all fundamental fields are quantized. On the other hand, this provides no explanation for the presence of other fields that are hypothesized in cosmology to explain observations and to make the different cosmological models work. For example, quantum gravity does not give any explanation for what could be the field that gives the accelerated phase hypothesized in the early universe, the {\em inflaton} \cite{Guth:1980zm}.

Some cosmological models postulate the existence of exotic matter, namely matter with equations for the energy that do not follow the ordinary ones. This serves to avoid the presence at the beginning of the expansion of the universe of a singularity, a point of infinite energy density that appears in the classical theory of gravity \cite{Tipler:1978}. This is not necessary considering the quantization of gravity: space-time quanta implies that such singularities cannot exist.
In a simple analogy, we can think that just as the electrons of an atom do not fall into its nucleus thanks to the quantization of energy, in the same way a gravitational collapse will not give rise to a singularity thanks to the quantization of space-time.

Note that in this presentation insofar dark energy was not mentioned. In fact, it is not necessary to think of dark energy as a form of matter. The reason is that the expansion of the universe, as far as we know today, is compatible with the cosmological constant that appears in the equations of General Relativity \cite{Bianchi:2010uw}. It is therefore an effect due to geometry, not to matter. It can be included in the quantum equations of gravity (in the context of Loop Quantum Gravity, this is not true in other contexts such as in the case of String Theories, where instead a positive cosmological constant is problematic).

A different consideration applies in the case of dark matter. This is, indeed, a big open question: Which matter only interacts gravitationally and yet constitutes most of the matter in our universe? Much research has been done to look for new particles that could explain it: unfortunately, to date, none of the particles hypothesized for this purpose have been found. An explanation in terms of some new form of matter becomes less and less credible. The alternative is to pay attention back to gravity. One possibility is to study modifications to Einstein's theory that explain all the observations on a galactic and cosmological scale. There are more or less successful attempts in this direction, even though the more we collect data to test General Relativity, the stronger the case insofar for no modifications.

One possibility, today among the favoured ones by physicists and which is simply based on a gravitational mechanism, is that dark matter may be simply black holes that formed in the primordial universe. This is today a very promising possibility, which could develop in different directions depending on what the mass (or mass distribution) of these primordial black holes is. There are several scenarios open today: both that of very large black holes, and that of small black holes. Unfortunately, making a direct detection of these black holes is very difficult: they are scattered and elusive! Their credibility as dark matter is therefore closely linked to the viability of the cosmological models in which they can be produced.

Finally, one possibility that I find very exciting is that dark matter could be composed not of black holes, but of remnants \cite{Rovelli:2018okm} left over after the ``death of black holes." This scenario is particularly interesting when compared with cosmological models where there is a bounce. The remnants, in this case, may have formed in the contraction phase of the universe and may have survived through the bouncing phase \cite{Rovelli:2018hba}.
How a black hole exhausts its black hole life and turns into a remnant depends on a quantum phenomenon of geometry. In this sense, black hole remnants are effectively a dark matter candidate that requires quantum gravity to be fully explained. The beauty of this model, I find, is that it requires no changes to physics other than what we know and what we expect (we expect gravity to be quantized).

\section{Conclusions}
To understand time, space, and matter in the primordial universe, we must give up relying solely on our common sense. 
This is what great science does: to show us beyond the veil, beyond the patterns that are useful to us for daily life but which may not be effective in capturing the essence of natural phenomena. Understanding the quantum characteristics of the space and time is requiring to build a new theory, a theory of quantum gravity.

To build a new theory we have to rely on philosophy. In fact, it is this that gives us conceptual tools to discern what, in the theories currently available to us, is essential, and what is instead an accident of the particular situation in which we study a given phenomenon. We must therefore proceed by studying the fundamentals, asking ourselves what concepts of space and time offer their most essential version. What concepts can we do without to describe space and time? Which ones are indispensable? It is through this path of conceptual cleaning that we put ourselves in the position of writing the new equations in a language compatible with both Quantum Mechanics and General Relativity.

Because of this need to rethink the basis, the recent research in cosmology and quantum gravity has given rise to an intense interactions between physicists and philosophers of science. On the one hand, philosophers of science study tentative theories with attention to their conceptual coherence and their incompleteness. On the other hand, physicists find inspiration not only in these analysis, but also in the traditional philosophical discussion on the nature of spacetime and matter. The classical philosophy, also in part inspired by science, has developed a variety of different manners for interpreting the notions of space and time. From Aristotle to Descartes, from Hume to Kant, and from Mach to Reichenbach, philosophers have provided a variety of different conceptualizations of the temporal and spacial structure of the world. Familiarity with this variety has become an indispensable tool for the advancement of cosmology and fundamental physics. For instance the notion of space as an entity, or as relations, or as an a priori condition for intuition, or as a tool to organize sensations or empirical data: all can be distinguished and each play a different role in a modern theory of gravity. The common sense and simple minded notion of space gets opened up into a complex set of concepts. 
Even richer is the construction of time, whose different properties are understood by different part of our science: from fundamental timeless quantum gravity to the general relativistic and Newtonian formulation, all the way down to thermodynamics and possibly even neurosciences.

Viceversa, this conceptual analysis is nourished and put to work by the scientific method,  by observations and experiments. We certainly do not yet have direct experiments proving the quantum properties of space-time (although, as I mentioned, these are now closer than previously imagined). This does not mean that we do not have any experimental data: we have a series of data that constraint the space of possible theories of quantum gravity. For example, we know that no micro black holes are produced at CERN. We also know that, if they ever exist, supersymmetric particles do not manifest themselves at the energies available at CERN or through astrophysical phenomena associated with dark matter. We know with great accuracy that the Lorentz symmetry is not violated. We know that gravitational waves travel at the speed of light. And so on: Nature is giving us clues to follow.\\[.1em]

Understanding space, time and matter in the early universe brings us to the edge of our knowledge: and it is there that philosophy and experience must meet.



\newpage
\begin{thebibliography}{1}

%
\bibitem{Vidotto:2015bza}
Francesca Vidotto,
\newblock  {Relational Quantum Cosmology.}
\newblock In {\em The Philosophy of Cosmology}, Khalil Chamcham, Joseph Silk, John D.  Barrow, and Simon Saunders eds.
\newblock  2017, {Cambridge University Press.}

\bibitem{Einstein:1917}
Albert Einstein.
\newblock  {Kosmologische Betrachtungen Zur Allgemeinen Relativit\"atstheorie.}
\newblock  {\em Preussische Akademie der Wissenschaften}, Sitzungsberichte, 1917.

\bibitem{Lemaitre:1927}
George Lema\^itre.
\newblock  {Un Univers Homog\`ene de Masse Constante et de Rayon Croissant Rendant Compte de la Vitesse Radiale des N\'ebuleuses Extra-Galactiques}
\newblock  {\em Annales de la Soci\'et\'e scientifique de Bruxelles}, A47: 49?59.

\bibitem{Einstein:1952}
Albert Einstein. 
\newblock  {Appendix V.} 
\newblock Added to A. Einstein (1952), {\em Relativity, the Special and the General Theory}, (pp. 135-157).

\bibitem{Hubble:1925}
Edward Hubble.
\newblock  {Cepheids in spiral nebulae.}
\newblock  {\em The Observatory}, 1925, Vol. 48, pages 139-142.

\bibitem{Smeenk:2012}
Chris Smeenk. 
\newblock  {Einstein?s role in the creation of relativistic cosmology.} 
\newblock In {\em Cambridge Companion to Einstein}, Michel Janssen and Christoph Lehner, eds.
\newblock 2012, Cambridge University Press.

\bibitem{Holton:1968}
Gerald Holton.
\newblock  {Mach, Einstein, and the Search for Reality}
\newblock  {\em Daedalus} 97 (1968) 636?673.

\bibitem{Rovelli:2021}
\newblock  Carlo Rovelli.
\newblock General Relativity: The Essentials.
\newblock 2021, Cambridge University Press.

%
\bibitem{Rovelli:2015gwa}
Carlo Rovelli.
\newblock The strange equation of quantum gravity.
\newblock {\em Class. Quant. Grav.} \textbf{32} (2015) no.12, 124005

\bibitem{WdW}
Bryce DeWitt.
\newblock Quantum Theory of Gravity. I. The Canonical Theory. 
\newblock {\em Phys. Rev.} 160 (1967) 5: 1113?1148.

%
\bibitem{Ashtekar:2007tv}
Abhay Ashtekar.
\newblock {An Introduction to Loop Quantum Gravity Through Cosmology}.
\newblock {\em Nuovo Cim.} B \textbf{122} (2007), 135-155.

\bibitem{Castro-Ruiz:2017gzs}
Esteban Castro-Ruiz, Flaminia Giacomini, {\v C}aslav Brukner
\newblock {Dynamics of quantum causal structures.}
\newblock {\em Phys. Rev.} X \textbf{8} (2018) no.1, 011047.

\bibitem{Hartle:1983ai}
Jim Hartle and Stephen Hawking,
\newblock {Wave Function of the Universe}.
\newblock {\em Phys. Rev.} D \textbf{28} (1983), 2960-2975

\bibitem{Albert:2000}
David Albert.
\newblock {Time and Chance}.
 \newblock 2000, Harvard University Press.

\bibitem{Rovelli:2014ssa}
Carlo Rovelli and Francesca Vidotto.
\newblock Covariant Loop Quantum Gravity: An Elementary Introduction to Quantum Gravity and Spinfoam Theory.
\newblock 2014, Cambridge University Press.

\bibitem{Gambini:2020ypi}
Rodolfo Gambini and Jorge Pullin.
\newblock  Loop Quantum Gravity for Everyone.
\newblock  2020, Oxford University Press.

\bibitem{Gambini:2011zz}
Rodolfo Gambini and Jorge Pullin.
\newblock  A first course in loop quantum gravity.
\newblock  2011, Oxford University Press.


\bibitem{Penrose:1966}
Roger Penrose.
\newblock  {Combinatorial quantum theory and quantized directions.}
\newblock  In {\em Advances in Twistor Theory}, L. P. Houghston, R. S. Ward eds., 1966.


\bibitem{Rovelli:2010ed}
Carlo Rovelli and Simone Speziale.
\newblock  {Lorentz covariance of loop quantum gravity}.
\newblock  {\em Phys. Rev.} D \textbf{83} (2011), 104029.

\bibitem{Ashtekar:2015dja}
Abhay Ashtekar and Aur\'elien Barrau,
\newblock  {Loop quantum cosmology: From pre-inflationary dynamics to observations}
\newblock  {\em Class. Quant. Grav.} \textbf{32} (2015) no.23, 234001.

\bibitem{Gozzini:2019nbo}
Francesco Gozzini and Francesca Vidotto,
\newblock  {Primordial Fluctuations From Quantum Gravity}.
\newblock  {\em Front. Astron. Space Sci.} \textbf{7} (2021), 629466.

\bibitem{Rovelli:2004tv}
\newblock  Carlo Rovelli.
\newblock Quantum gravity.
\newblock 2004, Cambridge University Press.

\bibitem{Guth:1980zm}
Alan Guth.
\newblock {The Inflationary Universe: A Possible Solution to the Horizon and Flatness Problems}.
\newblock {\em Phys. Rev. } D \textbf{23} (1981), 347-356.

\bibitem{Tipler:1978}
Frank Tipler.
\newblock {Energy conditions and spacetime singularities}.
\newblock {\em Phys. Rev.} D \textbf{17} (1978) 2521.

\bibitem{Bianchi:2010uw}
Eugenio Bianchi and Carlo Rovelli,
\newblock {Why all these prejudices against a constant?}
\newblock 2010 {\em preprint.} arXiv:1002.3966


\bibitem{Rovelli:2018okm}
Carlo Rovelli and Francesca Vidotto.
\newblock Small black/white hole stability and dark matter.
\newblock {\em Universe} \textbf{4} (2018) no.11, 127

\bibitem{Rovelli:2018hba}
Carlo Rovelli and Francesca Vidotto.
\newblock Pre-Big-Bang Black-Hole Remnants and Past Low Entropy.
\newblock {\em Universe} \textbf{4} (2018) no.11, 129


\end{thebibliography}
\end{document}